\documentclass[prd,twocolumn,superscriptaddress,showpacs,nofootinbib,preprintnumbers]{revtex4}

\usepackage{amsmath}
\usepackage{graphicx,bm}
\usepackage{dcolumn}    

\begin{document}
\title{Wave propagation in a weak gravitational field and the 
validity of the thin lens approximation}
\author{Teruaki Suyama}
\affiliation{Department of Physics, Kyoto University, Kyoto 606-8502, Japan}

\author{Ryuichi Takahashi}
\affiliation{Division of Theoretical Astrophysics, National Astronomical Observatory of Japan, Mitaka, Tokyo 181-8588,Japan}

\author{Shugo Michikoshi}
\affiliation{Department of Physics, Kyoto University, Kyoto 606-8502, Japan}

\preprint{KUNS-1969}

\begin{abstract}
Wave effects can be important for the gravitational 
lensing of gravitational waves.
In such a case, wave optics must be used in stead of
geometric optics.
We consider a plane wave entering a lens object and
solve numerically the wave equation for three lens
models: the uniform density sphere, the singular isothermal
sphere, and the Hernquist model.
By comparing our numerical solutions with the analytical solutions 
under the thin lens approximation, 
we evaluate the error of this approximation.
The results show that the relative error of the thin lens 
approximation is small if the geometrical thickness of 
the lens is much smaller than the distance between the lens
and the observer.
\end{abstract}


\maketitle

\section{Introduction}
Gravitational waves from coalescing compact binaries 
composed of neutron stars or black holes are the most
promising targets for ground-based as well as space-based
detectors. By applying the matched 
filtering technique which uses our theoretical predictions
of wave forms obtained by the post-Newtonian computations, 
we can extract the binary parameters such as the masses of 
each compact object, distance to the source, spatial positions
of the source and so on \cite{Cutler}.

One possibility which alters the predicted wave forms
calculated with high precision is the gravitational
lensing of gravitational waves.
If a massive object lies suitably between the source and the
observer, gravitational lensing of gravitational waves occurs.
One important point is that since the wavelength of gravitational
waves we are interested in is much larger than that of light,
a situation where the geometrical optics approximation breaks
down can be realized in some cases. As is discussed by many
authors \cite{Ohanian, Bliokh, Bontz, Thorne, Deguchi},
if the wavelength is larger than the Schwarzschild radius of
the lens object, the diffraction effect becomes important and
the magnification approaches to unity.
Therefore we must use wave optics rather than the geometric optics for
\begin{equation}
M_L \lesssim 10^8 M_{\odot} {\left( \frac{f}{\rm{mHz}} \right)}^{-1},
\label{int1} 
\end{equation}
where $ M_L $ is a mass of the lens and $ f $ is the frequency 
of the gravitational waves.
This frequency (mHz) is the case for the planned detector LISA~\cite{LISA}.

Further because the gravitational waves from 
a compact binary are coherent,
interference between lensed waves is important. 
Note that this situation is not in general realized 
in the case of gravitational lensing of electro-magnetic 
wave such as visible light. 
Since light is emitted from microscopic region
(usually atomic size) which is much smaller than the 
size of the source, each phase of the electro-magnetic wave
emitted from different points has no correlation and thus
interference effect vanishes.

If we assume the coalescence of SMBHs of mass 
$ 10^4 \sim 10^7 M_{\odot} $ 
as the source of the gravitational waves, it can be detected 
even if the sources are located at the cosmological distance
($ z > 5 $). Event rate of SMBH-SMBH merger for LISA is 
estimated as 
$ 0.1 \sim 10^2 ~{\rm event/yr} $ \cite{Haehnelt}
and lensing probability becomes several percent. 
Hence, some lensing events per year will be detected by LISA. 

Motivated by the fact that wave effects can be detected
for the gravitational lensing of gravitational waves,
there are now growing interests in the wave optics in
gravitational lensing 
\cite{Takahashi:2003ix,Seto:2003iw,Yamamoto:2003wg,Paolis,Nakamura:1997sw,Yamamoto,T.T.Nakamura,Baraldo:1999ny}. 
However we have to solve wave equation which is generally
partial differential equation between the source and the
observer in order to obtain the lensed wave form at the
observer. 
Except for a few special cases,
exact solutions of wave equation are not known at present.
Several authors have used thin lens approximation which 
reduces wave equation to double integral for
single lens object (for multi-lens objects, 
integration becomes multi-integral, 
see \cite{Yamamoto}) and thus makes the problem easier
 \cite{T.T.Nakamura,sef92}. 
In geometric optics the trajectory of light 
ray is obtained by solving geodesic equation, and
it is known that thin lens approximation is valid \cite{sef92}.
However there has been no studies or comments about 
the validity of the thin lens approximation in the 
framework of wave optics. 

In this paper, 
we develop a formulation to solve the wave equation
for a spherically symmetric lens,
where a partial differential equation reduces to a set
of ordinary differential equations.
We also solve those equations for simple lens models:
the uniform density sphere, the singular isothermal sphere
and the Hernquist model and evaluate the error of the
thin lens approximation.

This paper is organized as follows. 
In section II, 
we briefly review wave optics in gravitational 
lensing under the thin lens approximation. 
In section III, 
we develop a formalulation to solve the scattering problem of 
gravitational waves by a lens.
In section IV, 
we present our numerical results
and discuss the validity of the thin lens approximation.
Section V is a summary.
We use unit of $ c=1 $.

\section{Gravitationally Lensed Waveform under the Thin Lens Approximation}
We consider the wave propagation under the gravitational fields of a lens.
We assume that the spacetime metric is a Minkowski spacetime
plus a small perturbation due to the existence of a static
lensing object. Then the metric can be written as
\begin{equation}
ds^2 = g_{\mu \nu} dx^\mu dx^\nu = -(1+2U)dt^2+(1-2U)d {\vec x}^2,
\label{TLA1}
\end{equation}
where $ U $ is a Newtonian potential of the lensing object.
We consider a propagation of scalar waves $\phi$, 
instead of gravitational waves, 
since the wave equation for $\phi$ is the same as that for
gravitational waves \cite{Peters:1974gj}.
The scalar wave equation,
$ \partial_\mu ( \sqrt{-g} g^{\mu \nu} \partial_\nu \phi)=0 $, 
with the metric (\ref{TLA1}) is rewritten as
\begin{equation}
( \triangle + {\omega}^2 ) \phi(\vec{x}) = 4 {\omega}^2 U(\vec{x}) \phi (\vec{x}), \label{TLA2}
\end{equation}
where we assume that the wave is monochromatic with the angular frequency
 $ \omega $.
The above equation was solved by using the Kirchhoff diffraction
 integral (see \cite{sef92}, Sec.4.7 and 7) under the thin lens 
approximation.

\begin{figure}[t]
  \includegraphics[width=6cm,clip]{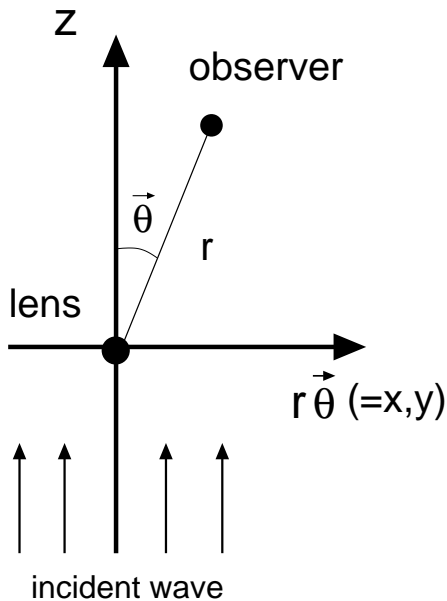}
\caption{Lensing configuration. The lens is the origin of
 coordinate axes, while the observer position is $(r, {\vec {\theta}})$
 with $\theta \ll 1$.
The incident wave is a plane wave propagating in the z-direction. 
}
\label{fthin} 
\end{figure}

We show lensing configuration in Fig.\ref{fthin}.
We choose a position of the lens as the origin of the polar coordinate 
 system ($r,\theta,\phi$).
The observer position is $r$ and ${\vec \theta}=(\theta \cos \phi,
 \theta \sin \phi)$ with $\theta \ll 1$.
The incident wave is a plane wave propagating in the z-direction.
Denoting the incident wave as $\phi^0$, we have
 $\phi^0=e^{i \omega r \cos \theta}$.

In this section, we assume the thin lens approximation, in which the wave is
 scattered only on the thin lens plane
 at $z=0$, and the lens is characterized by the surface mass density
 $\Sigma({\vec{s}})$, where $\vec{s}=(x,y)$.
The two dimensional potential $\psi (\vec{s})$ in $z=0$ plane is defined as
\begin{equation}
\psi (\vec{s}) = 2 \int_{-\infty}^{\infty} dz U(\vec{s},z). \label{TLA5}
\end{equation}
Here, $\psi$ is also obtained from the surface density
 by using $\nabla^2_s \psi(\vec{s}) = 8 \pi \Sigma(\vec{s})$. 

It is useful to define the amplification factor $ F $ (which is called the 
transmission factor in Ref.\cite{sef92}) as $ F = \phi/{\phi}_o $, where
 $\phi$ is the gravitationally lensed waveform obtained by solving
 Eq.(\ref{TLA2}) and $\phi^0$ is the incident wave.
The amplification factor at the observer under the thin lens approximation
 is given by \cite{sef92}
\begin{equation}
F_{thin}(r, {\vec \theta})=\frac{\omega}{2 \pi i r} \int d^2 s
 e^{i \frac{\omega}{2 r}{| r \vec{\theta} - \vec{s} |}^2-i \omega \psi
 (\vec{s})} \label{TLA6}.
\end{equation}
Here, $F_{thin}$ is normalized so that
$F_{thin}=1$ for $ \psi=0 $.

\section{Formulation of Numerical Calculation}
In this section, we develop a formulation to solve the scattering 
problem of gravitational waves by lensing object which
is applicable to the case where the lens potential is
spherically symmetric.
A situation we will consider is 
that plane wave is entering weak gravitational field 
which is spherically symmetric. 
When the lensing object is spherically symmetric, 
a scattering problem can be reduced to a problem of 
determining so-called phase shift which is used to
probe the nature of nuclear physics and is also useful 
for the scattering by BHs where the assumption 
of weak gravitational field breaks down \cite{Andersson:2000tf}.

We have to solve Eq.~(\ref{TLA2}) in order to evaluate 
quantities such as the amplification factor which can be 
compared with the one derived under the thin lens approximation.
To solve Eq.~(\ref{TLA2}), we choose a center of the lens as the origin of 
coordinate system. Then $ U $ becomes a function which depends only on the
 radius coordinate $ r $. 
Now let us expand $ \phi $ in terms of Legendre function
\begin{equation}
\phi (\xi, \theta) = \sum\displaylimits_{\ell=0}^{\infty} \frac{g_{\ell}(\xi)}{\xi} P_{\ell} (\cos \theta), \label{formu3}
\end{equation}
where $ \xi $ is a dimensionless variable defined by $ \xi \equiv r \omega $. 
Then equations for $ g_{\ell} (\xi) $ are
\begin{equation}
\left( \frac{d^2}{d {\xi}^2} +1-4 {\tilde U}(\xi) - \frac{\ell (\ell+1)}{ {\xi}^2 } \right) g_{\ell} (\xi) =0, \label{formu4}
\end{equation}
where we have used $ {\tilde U} $ in stead of $ U $ in order to stress that $ {\tilde U} $ 
is a function of $ \xi $. 

For a point mass lens, i.e. $ {\tilde U} (\xi) = -p/(2 \xi) $, where 
$ p \equiv 2GM \omega $ and $ M $ is a lens mass, analytic solutions of Eq. (\ref{formu4}) 
are known as Coulomb wave functions \cite{Abramowitz}. The solution which is regular at $ \xi = 0 $   
is
\begin{eqnarray}
F_{\ell}(-p, \xi) = e^{\frac{\pi}{2}p} \frac{\Gamma (\ell+1-ip)}{\Gamma (2(\ell+1))} 2^{\ell} {\xi}^{\ell+1}
e^{i \xi+i {\sigma}_{\ell}} \nonumber \\
\times F(\ell+1-ip; 2(\ell+1); -2 i \xi), \nonumber \\
{\sigma}_{\ell} \equiv \arg \Gamma (\ell+1+ip) , \label{formu5}
\end{eqnarray}
where $ F $ is the confluent hypergeometric function. The solution which is singular at $ \xi =0 $ is
\begin{eqnarray}
G_{\ell} (-p,\xi) &=& {\xi}^{\ell+1} 2^{\ell} e^{-\frac{\pi}{2}p}
 {(-i)}^{2 \ell+1} e^{i \xi-i {\sigma}_{\ell}} \nonumber \\
 && \times U(\ell+1-ip; 2(\ell+1); -2i \xi) +c.c., \label{formu6}
\end{eqnarray}
where $ U $ is defined as
\begin{equation}
U(a; b; z) =\frac{1}{\Gamma (a)} \int_0^{\infty} dt e^{-z t} t^{a-1} {(1+t)}^{b-a-1}. \label{formu7}
\end{equation}
The asymptotic form of these functions are 
\begin{eqnarray}
F_{\ell} (-p, \xi) \xrightarrow[\xi \gg 1]{} \sin (\xi +p \log 2\xi -\frac{\pi}{2} \ell - {\sigma}_{\ell} ), \nonumber \\
G_{\ell} (-p, \xi) \xrightarrow[\xi \gg 1]{} \cos (\xi +p \log 2\xi -\frac{\pi}{2} \ell - {\sigma}_{\ell} ). \label{formu7.5}
\end{eqnarray}
The term $ p \log 2 \xi $ in the phase of trigonometric functions represents the nature of 
long range force which is characteristic of Coulomb force.

On the contrary, a solution of Eq. (\ref{TLA2}) that a plane wave is entering a
point mass lens is well known and is given by \cite{Messiah} 
\begin{equation}
{\phi}_p = e^{\frac{\pi}{2} p} \Gamma (1-ip) e^{i \xi \cos \theta} F(ip; 1; i \xi (1- \cos \theta)). \label{formu8}
\end{equation}
Because the solution Eq. (\ref{formu8}) is regular at $ \xi = 0 $,
 it is written as
a partial wave sum of regular Coulomb wave function,
\begin{eqnarray}
{\phi}_p = \sum\displaylimits_{\ell=0}^{\infty} a_{\ell} \frac{F_{\ell}(-p, \xi)}{\xi} P_{\ell} (\cos \theta), \nonumber \\
a_{\ell} = i^{\ell} (2\ell+1) e^{-i {\sigma}_{\ell}}. \label{formu9}
\end{eqnarray}

For an extended lensing object, analytic solution of Eq. (\ref{TLA2}) 
does not exist usually. 
However if the lensing object exists only in a 
finite region, 
then the solution of Eq. (\ref{TLA2}) outside the lens 
can be written as a summation of partial waves which are now 
a linear combination of two independent Coulomb wave functions. 
By determining a coefficient of
each Coulomb wave function, we can calculate the wave form $ \phi $ outside 
the lens. In the aim of only determining the wave form far from the lensing 
object, we don't need to know the expression of $ \phi $ inside the lens.

Now let us write the solution of Eq. (\ref{TLA2}) as
\begin{equation}
\phi = {\phi}_p+{\phi}_s. \label{formu10}
\end{equation}
Thus $ {\phi}_s $ represents the scattered wave which 
arises due to the deviation of lens from a point mass. 
There may be no incoming scattered wave to the lens 
from infinity, so we assume the asymptotic form of $ {\phi}_s $ as
\begin{equation}
{\phi}_s (\xi, \theta) = \sum\displaylimits_{\ell=0}^{\infty} \frac{e^{i \xi +ip \log 2 \xi -i \frac{\pi}{2}\ell -i {\sigma}_{\ell}}}{2i \xi} s_{\ell} P_{\ell} (\cos \theta), \label{formu11}
\end{equation}
where $ s_{\ell} $ are undetermined complex numbers, 
but not arbitrary. 
In order that $ \phi $ in Eq. (\ref{formu10})
satisfy the wave equation (\ref{TLA2}), 
$ s_{\ell} $ must be related to $ a_{\ell} $ as
\begin{equation}
s_{\ell} = a_{\ell} (e^{2i {\delta}_{\ell}}-1), \label{formu12}
\end{equation}
where $ {\delta}_{\ell} $ are real numbers. 
In terms of $ {\delta}_{\ell} $, $ \phi $ is written as
\begin{eqnarray}
\phi (\xi, \theta) = \sum\displaylimits_{\ell=0}^{\infty} a_{\ell} e^{i {\delta}_{\ell}} 
\left( \cos {\delta}_{\ell} \frac{F_{\ell}(-p,\xi)}{\xi} \right. \nonumber \\
{} \left. +\sin {\delta}_{\ell} \frac{G_{\ell}(-p,\xi)}{\xi} \right) P_{\ell} (\cos \theta). \label{formu13} 
\end{eqnarray}
Thus we can calculate wave form outside the lens object
by determining the phase shift $ {\delta}_{\ell} $.
Here, ${\delta}_{\ell}$ are determined by matching a solution of 
Eq.~(\ref{formu4}) with Eq. (\ref{formu13}) at a radius
 $\xi_0$ being outside the lens.
Then, we have
\begin{equation}
\tan {\delta}_{\ell} = \frac{ g_{\ell} ({\xi}_o) F'_{\ell} (-p,{\xi}_o) -g'_{\ell} ({\xi}_o) F_{\ell}(-p,{\xi}_o)}{g'_{\ell}({\xi}_o) G_{\ell}(-p,{\xi}_o)-g_{\ell}({\xi}_o) G'_{\ell}(-p,{\xi}_o)}, \label{formu14}
\end{equation}
where $ g_{\ell} $ is a solution of Eq.~(\ref{formu4})
which is to be calculated numerically.
We set the initial condition of Eq.~(\ref{formu4}) 
at $ \xi = 0 $ is that $ g_{\ell} $ is regular.
The range of $ {\delta}_{\ell} $ is from
$ -{\pi}/{2} $ to $ {\pi}/{2} $.

\section{Results}
We investigate the validity of the thin lens approximation for
three lens models; the uniform density sphere, the singular isothermal
sphere and the Hernquist model.

\subsection{Uniform density sphere}
We first present the results for uniform density sphere
which has the simplest structure next to the point mass.
The gravitational potential for the uniform
density sphere is given by
\begin{eqnarray}
U(r) = \left\{
\begin{array}{@{\,}ll}
-\frac{GM}{2R} \left(3-\frac{r^2}{R^2} \right) & \ \ \
 \mbox{($ r \le  R $)} \\
-\frac{GM}{r} & \ \ \  \mbox{($ r \ge R $)}, 
\end{array}
\right. \label{ana11}
\end{eqnarray}
where $r$ is the distance from the center of the
sphere, $R$ is the radius of the sphere and $ M $ is
the lens mass.
Here we consider the case that $ R $ is larger than the 
Einstein radius $ r_E $, in which case the effect of
the size of the lens object is expected to be
important.
For $ R < r_E $, the result is almost the same as for point 
lens mass and it is known that amplitude of the 
amplification factor for the point mass lens coincides
with that in the thin lens approximation \cite{T.T.Nakamura}.

\begin{figure}[t]
  \includegraphics[width=7cm,clip]{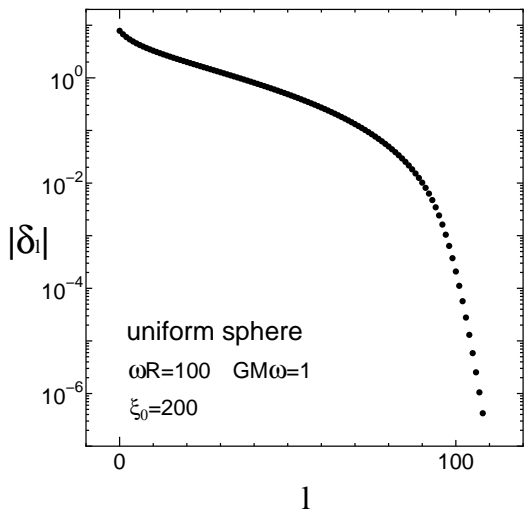}
\caption{$ {\delta}_{\ell} $ as a function of $ \ell $ for the
uniform density sphere.}
\label{fig2} 
\end{figure}

\begin{figure*}[t]
  \includegraphics[width=13cm,clip]{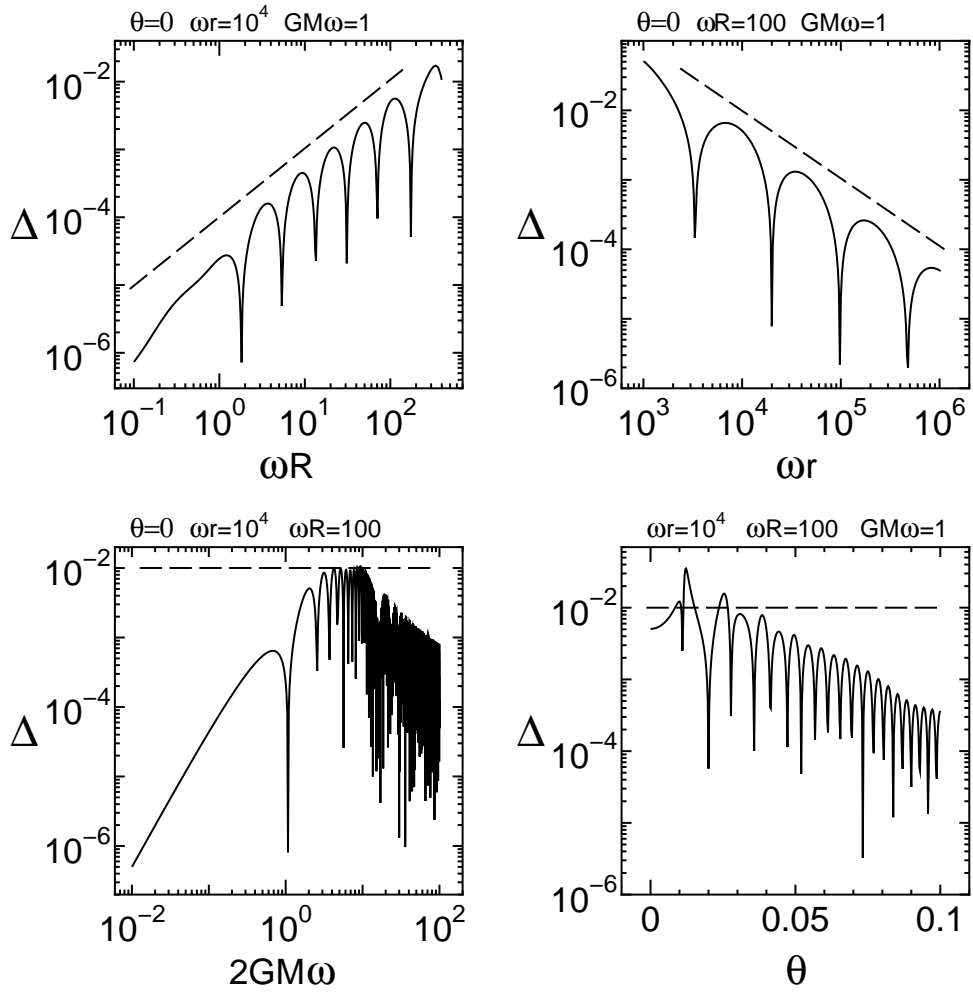}
\caption{Error of the amplification factor for the uniform density
sphere.
The horizontal axis is the radius $\omega R$ (top left), the distance
 $\omega r$ (top right), the mass $2 G M \omega (=p)$ (bottom left),
 and the angle $\theta$ (bottom right). 
The dashed lines denote the ratio of radius to distance $R/r$.
}
\label{fig3} 
\end{figure*}

The amplification factor in the thin lens approximation
Eq.~(\ref{TLA6}) for the case of uniform density sphere
is written as
\begin{equation}
F_{thin} (r,\theta) = -i \frac{\omega}{r} e^{\frac{i}{2} \omega r
 \theta^2} \int_0^{\infty} ds s J_0 (\omega s \theta) e^{iw (
 \frac{s^2}{2 r}-\psi (s) )}, \label{form15}
\end{equation}
where $ J_0 $ is the 0-th Bessel function and $ \psi $ is
given by
\begin{widetext}
\begin{eqnarray}
\psi (s) = \left\{
\begin{array}{@{\,}ll}
4M \log \left[ 1+\sqrt{1-(s/R)^2} \right]
 -\frac{4 M}{3} \left(4-(s/R)^2 \right)
 \sqrt{1-(s/R)^2} & \ \ \  \mbox{($ s \le  R $)} \\
4M \log (s/R) & \ \ \  \mbox{($ s \ge R $)}. 
\end{array}
\right. \label{form16}
\end{eqnarray}
\end{widetext}

Next let us calculate $ F $ numerically by using the
method developed in the previous section.
We first show $ {\delta}_{\ell} $ as a function of $ \ell $
in fig.~\ref{fig2}.
We see that $ {\delta}_{\ell} $ decreases as $ \ell $
increases.
In particular, above around $ \ell =90 $,
$ {\delta}_{\ell} $ rapidly approaches zero.
This is because the $ \ell $ -th partial wave can be interpreted
as an incident particle with impact parameter $ \ell/ \omega $.
Partial waves of $ \ell \gtrsim \omega R $ pass through the
lens potential outside the lens object and the gravitational
effects on these partial waves are the same as point mass,
which implies $ {\delta}_{\ell} $ becomes zero.

Fig.\ref{fig3} shows the error of the amplification factor  
for the uniform density sphere as a function
of lens parameters 
$ \omega R, \omega r, GM \omega $ and $ \theta $ respectively. 
The vertical axis is the error $\Delta$ defined as, 
$ \Delta^2 \equiv (|F|-|F_{thin}|)^2 / |F_{thin}|^2 $.
The normalized Einstein radius is 
$\omega r_E = 2 \omega \sqrt{G M r} = 200 (GM \omega/1)^{1/2} (\omega r/10^4)^{1/2}$.
The radius of sphere $\omega R (=100)$ 
is comparable to the Einstein radius.

The top left and right panels show $ \Delta $ as a
function of $ \omega R $ and $ \omega r $ respectively.
We find that averaged in $ R $ or $ r $ over a 
period of oscillation, 
$ \Delta $ is proportional to $ R/r $.
Also we find that $ \Delta $ remains 
smaller than $ R/r $ which is much
smaller than unity.
(The dashed lines denote the ratio 
of the radius to distance $ R/r $.)
This suggests that geometrical thickness of 
lens $ R/r $ is a suitable measure of the 
validity of thin lens approximation in wave
optics.
 
The bottom left panel shows dependence of
$ \Delta $ on $ GM \omega $.
We find that $ \Delta $ has a peak around 
$ GM \omega =10 $ and decreases as 
$ GM \omega $ increases for
$ GM \omega >10 $.
This result shows that the thin lens 
approximation is valid for the wavelength
where we can use geometric optics instead 
of wave optics.
The reason why thin lens approximation 
becomes valid in the geometric optics is that
the deflection angle can be evaluated by using the thin lens potential
if the distance between lens and observer
is much larger than the lens size (see Ref.\cite{sef92}, p.124).
For this panel too, $ \Delta $ remains smaller
than $ R/r $ in all range of frequencies of
calculation.

The bottom right panel shows dependence of
$ \Delta $ on $ \theta $.
We find that $ \Delta $ becomes maximum at 
close to $ \theta =0 $ and decreases as $ \theta $ 
becomes larger.
This is because for large $ \theta $ such that
$ r \theta \gtrsim R $ 
($ \theta \gtrsim 0.01 $ for a case of fig.~\ref{fig3}),
the size effect of lens becomes negligible and
the lensed waveform becomes the same as for
the point mass lens for which it is known that 
thin lens approximation is valid.
Except for a small region of $ \theta $ around 
where $ \Delta $ takes maximum value, 
$ \Delta $ is smaller than $ R/r $.

To summarize, 
for the uniform density sphere, 
the relative error of the amplification factor in the 
thin lens approximation is suppressed within the ratio 
of its radius to distance $R/r$, 
which is much smaller than unity in a realistic 
astrophysical situations and is the largest
for $ \lambda \sim GM $ ($ \lambda $ is the wavelength).

\subsection{Singular Isothermal Sphere (SIS) model}
We also calculated $ \Delta $ for singular 
isothermal sphere which is a model of galaxies,
dark matter haloes and star clusters.
The density profile of SIS model is 
$\rho(r) = v^2/ (2 \pi r^2)$ where
$v$ is the velocity dispersion. 
For numerical calculation, 
we have to introduce cutoff radius $r_c$ 
because gravitational potential does not 
approaches to $ \propto r^{-1} $ far from
the lens without cutoff.
Thus we here assume that the density vanishes for $r > r_c$.
Newton potential becomes
\begin{eqnarray}
U(r) = \left\{
\begin{array}{@{\,}ll}
 \frac{GM}{r_c} \left( \ln \frac{r}{r_c} -1 \right) &
 \ \ \ \mbox{($ r \le  r_c $)} \\
-\frac{GM}{r} & \ \ \  \mbox{($ r \ge r_c $)}, 
\end{array}
\right. \label{sis1}
\end{eqnarray}
where $M$ is the mass inside the cutoff radius $r_c$ :
$M = 2 v^2 r_c $.

$F_{thin}$ is given by Eq.(\ref{form15}), 
and $\psi(s)$ is given by 
\begin{widetext}
\begin{eqnarray}
\psi(s) = \left\{
\begin{array}{@{\,}ll}
 - \frac{8GM}{r_c} \sqrt{r_c^2-s^2} + \frac{4GM}{r_c} s {\mbox{arctan}}
 \frac {\sqrt{r_c^2-s^2}}{s} + 4GM \ln \left( 1+ \sqrt{1-(s/r_s)^2} \right),
 & \ \ \ \mbox{($ s \le  r_c $)} \\
 8GM \ln (s/r_c)
 & \ \ \ \mbox{($ s \ge r_c $)}
\end{array}
\right. 
\end{eqnarray}
\end{widetext}

\begin{figure*}[t]
  \includegraphics[width=13cm,clip]{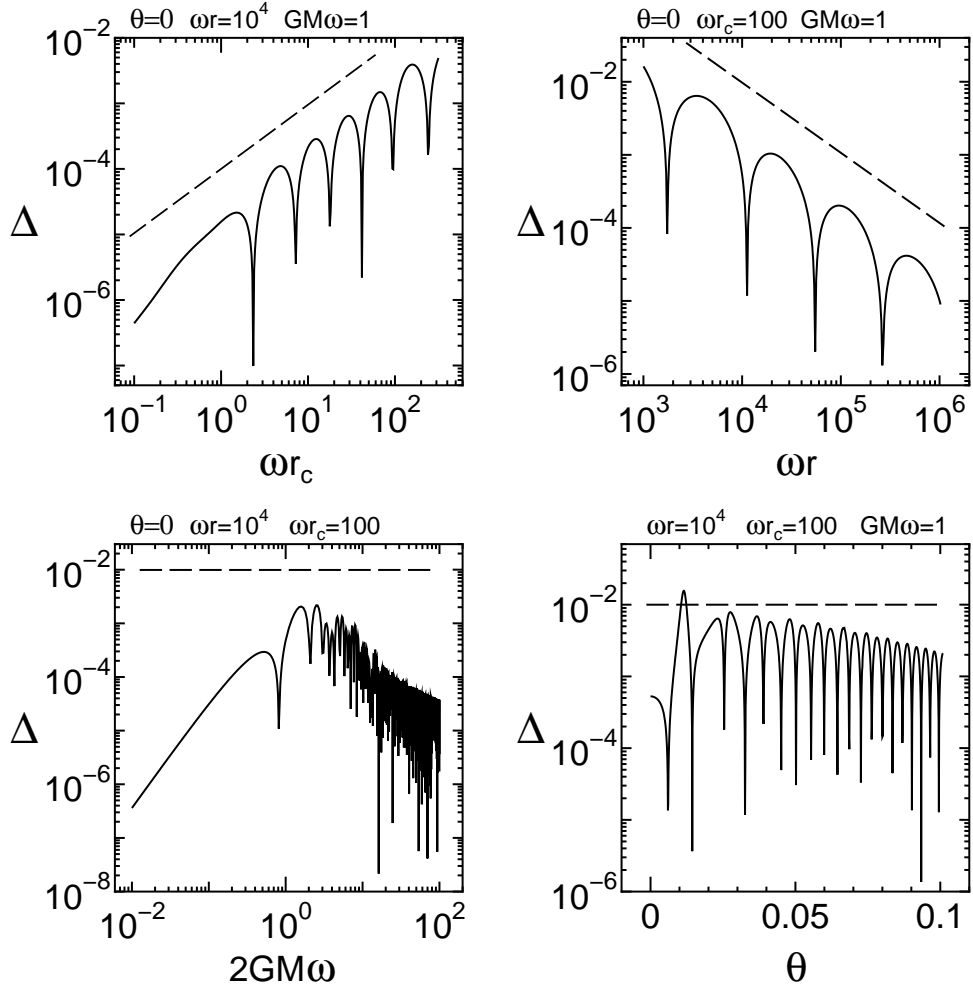}
\caption{Same as Fig.\ref{fig3}, but for the SIS model.
The dashed lines denote the ratio of cutoff radius to distance $r_c/r$.
}
\label{fig4} 
\end{figure*}

Fig.~\ref{fig4} presents dependence of 
$ \Delta $ on parameters, 
$ \omega r_c $, $ \omega r $, $ \omega GM $ and $ \theta $.
We find that the qualitative behavior is the same as
the case of uniform density sphere.
We see from the top left panel and the top right panel 
that thin lens approximation becomes worse as cutoff
approaches to observer's distance $ r $ from the center
of SIS.

\begin{figure*}[t]
  \includegraphics[width=13cm,clip]{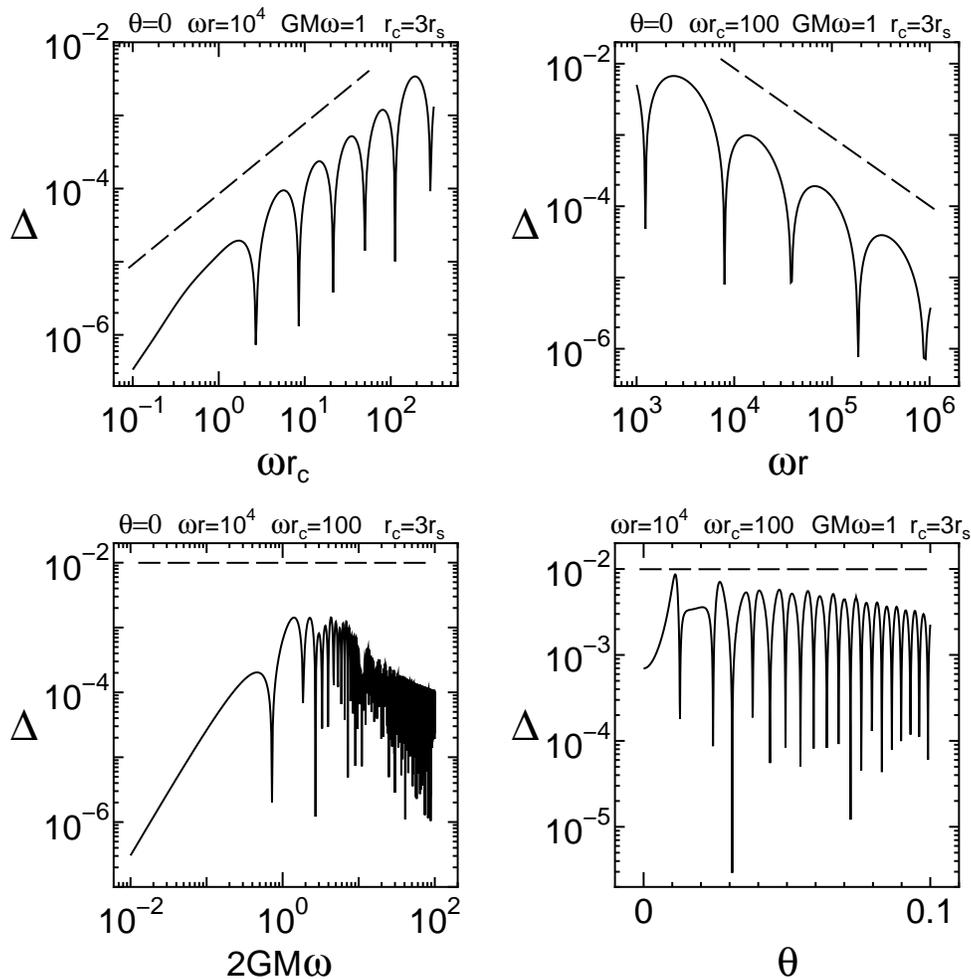}
\caption{Same as Fig.\ref{fig3}, but for the Hernquist model.
The dashed lines denote the ratio of cutoff radius to distance $r_c/r$.
}
\label{fig5} 
\end{figure*}

\subsection{Hernquist model}
We also calculated $ \Delta $ for 
Hernquist model \cite{Hernquist:1990be}, 
which fits well the luminosity distribution 
of many elliptical galaxies and bulges. 
Its density profile is given by
\begin{equation}
\rho (r) = \frac{ {\rho}_s }{ (r/r_s) {(1+r/r_s)}^3}, \label{hern1}
\end{equation}
where $ r_s $ is a scale length and $ {\rho}_s $ is a 
characteristic density.
For numerical calculation, 
we introduce cutoff radius $r_c$ for the same
reason as SIS.
For $r<r_c$ the density is given in Eq.(\ref{hern1}), 
while for $r>r_c$ the density vanishes.
Then, Newton potential becomes
\begin{eqnarray}
U(r) = \left\{
\begin{array}{@{\,}ll}
-\frac{GM}{r+r_s} \left( \frac{r_s+r_c}{r_c} \right)^2 + \frac{r_s}{r_c} GM
 & \ \ \ \mbox{($ r \le  r_c $)} \\
-\frac{GM}{r} & \ \ \  \mbox{($ r \ge r_c $)}, 
\end{array}
\right. \label{hern}
\end{eqnarray}
where $M$ is the mass inside the cutoff radius $r_c$ :
$M = 2 \pi \rho_s r_s^3 [r_c/(r_s+r_c)]^2 $.

$F_{thin}$ is given by Eq.(\ref{form15}), 
and $\psi(s)$ is given by 
\begin{widetext}
\begin{eqnarray}
\psi(s) = \left\{
\begin{array}{@{\,}lll}
 -4 M \frac{r_s^2+2 r_s r_c}{r_c^2} {\mbox{arccosh}} \frac{r_s}{s} \
 + 4 M \ln s + 4 \frac{r_s}{r_c^2} M \sqrt{r_s^2-s^2} +
 \frac{4 M r_s (r_s+r_c)^2}{r_c^2 \sqrt{r_s^2-s^2}}
 {\mbox{arctanh}} \left[ \frac{\sqrt{(r_s^2-s^2)(r_c^2-s^2)}}{s^2+r_c r_s}
 \right],
 & \ \ \ \mbox{($ s \le  r_s $)} \\
 -4 M \frac{r_s^2+2 r_s r_c}{r_c^2} {\mbox{arccosh}} \frac{r_s}{s} \
 + 4 M \ln s + 4 \frac{r_s}{r_c^2} M \sqrt{r_s^2-s^2} +
 \frac{4 M r_s (r_s+r_c)^2}{r_c^2 \sqrt{s^2-r_s^2}}
 {\mbox{arctan}} \left[ \frac{\sqrt{(s^2-r_s^2)(r_c^2-s^2)}}{s^2+r_c r_s}
 \right], 
 & \ \ \ \mbox{($ r_s \le s \le  r_c $)} \\
 4M \ln s. & \ \ \  \mbox{($ r \ge r_c $)}, 
\end{array}
\right. 
\end{eqnarray}
\end{widetext}

Fig.~\ref{fig5} shows $ \Delta $ for Hernquist model.
We see that the behavior of $ \Delta $ is almost the
same as SIS.

\section{summary}
In this paper we discussed the validity of the 
thin lens approximation in the framework of wave
optics.

In sec III,
we developed a formalism to solve the wave equation for
the spherically symmetric potential.
In this case, 
the partial differential equation can be reduced to 
a set of ordinary differential equations.
The method we used is to determine the so-called
phase-shift which represents the difference
of scattered waves between the point mass
lens and an arbitrary spherically symmetric potential.
This formalism is only applicable to the spherically
symmetric lens whose size is finite.
For lens models such as the SIS profile which
extends to infinity, 
we have to introduce cutoff to make the total 
mass finite.

We also solved the wave equation numerically for the 
spherically symmetric potential.
By numerical calculations, 
we found that the error of the thin lens 
approximation for the simple lens models 
is the same as or smaller than the geometric 
thickness of the lens, $ s/r $,
where $ s $ is the size of the lens and $ r $ is
the distance between the lens and the observer.
The error is the largest for the wavelength
comparable to the Schwarzschild radius
of the lens.

\acknowledgements

We would like to thank Takahiro Tanaka for useful comments and
encouragement. 
We would also like to thank Takashi Nakamura, Misao Sasaki, and 
Atsushi Taruya for useful comments.

\end{document}